\documentclass[prl,twocolumn,showpacs,superscriptaddress]{revtex4-1}
\usepackage{graphicx}
\usepackage[dvips]{color}
\usepackage{amssymb}

\begin{document}
\title{Polar phase of superfluid $^3$He in anisotropic aerogel}

\author{V.\,V.\,Dmitriev}
\email{dmitriev@kapitza.ras.ru}
\affiliation{P.L.~Kapitza Institute for physical problems of RAS, 119334, Moscow, Russia}
\author{A.\,A.\,Senin}
\affiliation{P.L.~Kapitza Institute for physical problems of RAS, 119334, Moscow, Russia}
\author{A.\,A.\,Soldatov}
\affiliation{P.L.~Kapitza Institute for physical problems of RAS, 119334, Moscow, Russia}
\affiliation{Moscow Institute of Physics and Technology, 141700, Dolgoprudny, Russia}
\author{A.\,N.\,Yudin}
\affiliation{P.L.~Kapitza Institute for physical problems of RAS, 119334, Moscow, Russia}

\date{\today}

\begin{abstract}
We report the observation of a new superfluid phase of $^3$He -- polar phase. This phase appears in
$\bf^3$He confined in a new type of aerogel with nearly parallel
arrangement of strands which play a role of ordered impurities.
Our observations qualitatively agree with theoretical predictions and
suggest that in other systems with unconventional Cooper pairing
(e.g. in unconventional superconductors) similar phenomena may be found in presence of anisotropic impurities.
\end{abstract}

\pacs{67.30.hm, 67.30.er, 67.30.hj, 74.20.Rp}
\maketitle

{\it Introduction.---}One of examples of superfluid Fermi systems with unconventional Cooper pairing is superfluid $^3$He where the pairing occurs with spin and orbital angular momentum equal 1. In isotropic space the free energy and the superfluid transition temperature are degenerate with respect to spin and orbital momentum projections. This allows a variety of superfluid phases with the same transition temperature, but in zero magnetic field only two phases (A and B) with the lowest energy are realized in bulk $^3$He \cite{VW,ORL,Gr73}. An anisotropy of the space may lift the degeneracy and other phases can be stabilized. So, in presence of magnetic field the degeneracy with respect to spin is lifted and the A$_1$ phase appears in a narrow region below the superfluid transition temperature \cite{a1}. As shown in \cite{AI} and in further theoretical works \cite{Sauls,Fom,Ik}, the degeneracy with respect to orbital momentum projections may be lifted by anisotropic impurities, e.g.
by globally anisotropic aerogel which strands are aligned on average along the same direction (``stretching'' anisotropy). In this case a new superfluid phase of $^3$He, the polar phase, may be stabilized below the transition temperature. It was predicted that on further cooling the 2nd order transition into a polar distorted A phase should occur and the distortion should decrease with cooling. At lower temperatures the 1st order transition into a polar distorted B phase was also expected.

In most of experiments with $^3$He in aerogel the samples of silica aerogel are used. They consist of chaotically distributed $SiO_2$ strands with diameters of $\sim$3~nm and with the average separation of $\sim$100~nm. The superfluid coherence length at different pressures is in the range of 20$\div$80~nm, so the strands play a role of impurities. It is established \cite{Parp,Halp0} that the superfluid transition temperature of $^3$He in aerogel ($T_{ca}$) is lower than the transition temperature ($T_c$) of bulk $^3$He but order parameters of observed superfluid A-like and B-like phases correspond to those of A and B phases of bulk $^3$He \cite{Osh,weB,we4,HalpA}. Global anisotropy of silica aerogel may influence the superfluid phase diagram but, although some unclear behavior is observed \cite{halp15}, no evidence of the polar distortion in A-like and B-like phases is found. Much larger ``stretching'' anisotropy is inherent to ``nematically ordered'' aerogel (N-aerogel) which strands are oriented along the same direction $\bf \hat \zeta$. There are two types of N-aerogel: ``Obninsk aerogel'' and nafen \cite{Ask2}. In recent experiments with $^3$He in ``Obninsk aerogel''  two superfluid phases were observed: the polar distorted A phase at higher temperatures and the polar distorted B phase at lower temperatures \cite{we1,we2,we3}. These observations qualitatively agree with theoretical predictions, however, the existence of the polar phase has not been proved.

Here we, for the first time, have investigated superfluid $^3$He confined in nafen, which has much larger
overall density than ``Obninsk aerogels''. We have found that in $^3$He in nafen the superfluid transition occurs into the pure polar phase. We also have carried out additional experiments which show that in ``Obninsk aerogel'' we obtain the polar distorted A phase but the polar phase is still not realized.

\begin{figure}[!t]
\centerline{\includegraphics[width=0.9\columnwidth]{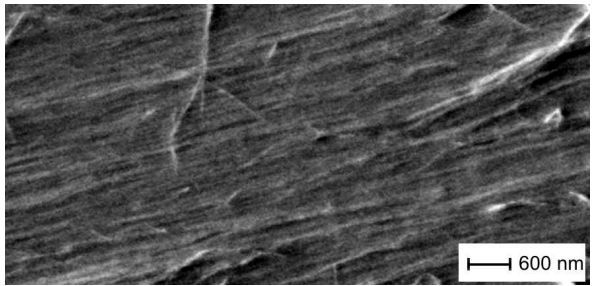}}
\caption{The SEM image of the surface of nafen-90.}
\label{f1}
\end{figure}

{\it Samples and Methods.---} We have used 2 samples of nafen and 2 samples of ``Obninsk aerogel'' which have a shape of cuboid with sizes of $\sim$4~mm.  Experimental chambers were similar to those described in \cite{we1}.
Nafen samples were produced by ANF Technology Ltd (Tallinn, Estonia). They consist of $Al_2O_3$ strands and have porosities of 97.8\% (the sample ``nafen-90'' with the overall density of 90~mg/cm$^3$) and 93.9\% (``nafen-243'' with the density of 243~mg/cm$^3$). Samples of ``Obninsk aerogel'' consist of $AlOOH$ strands and were produced in Leypunsky Institute (Obninsk, Russia). Their porosities are 97.9\% (the sample ``Obninsk-50'' with the density of 50~mg/cm$^3$) and 99.6\% (``Obninsk-8'' with the density of 8~mg/cm$^3$). Strands of all samples are nearly parallel to one another (Fig.~\ref{f1}) and have diameters 6$\div$9~nm. For more information about the structure of the samples see \cite{Ask2}.
In the limit of zero temperature the spin diffusion of $^3$He in N-aerogel is anisotropic
and the ratio of spin diffusion coefficients along and transverse to the strands is an important parameter of the theory.
For Obninsk-8 this ratio is $\sim$1.5 \cite{diff0}, while for nafen-90 and nafen-243 it equals $\sim$3.3 and $\sim$8.1 respectively \cite{diff}.

Experiments were performed using continuous wave (CW) and pulse NMR in magnetic fields 10$\div$37~mT (corresponding NMR frequencies are 330$\div$1200~kHz) and at pressures 0.2$\div$29.3~bar. We were able to rotate the external steady magnetic field {\bf H} by an arbitrary angle $\mu$ with respect to $\bf \hat \zeta$. Necessary temperatures were obtained by a nuclear demagnetization cryostat and were measured by a quartz tuning fork calibrated by measurements of the Leggett frequency in bulk $^3$He-B. To avoid a paramagnetic signal from surface solid $^3$He, the samples were preplated by $\sim$2.5 atomic monolayers of $^4$He.

{\it Spin dynamics in polar distorted A phase of $^3$He in N-aerogel.---}
The general form of the order parameter of polar, polar distorted A and pure A phases is:
\begin{equation}
\label{A}
A_{{\nu}k} =\Delta_0 e^{i\varphi}d_{\nu}\left(am_k+ibn_k\right),
\end{equation}
where $\Delta_0$ is the gap parameter, $e^{i\varphi}$ is the phase factor, ${\bf d}$ is the unit spin vector, ${\bf m}$ and
${\bf n}$ are mutually orthogonal unit vectors in orbital space and $a^2+b^2=1$. For the A phase $a=b$, for the polar distorted
A phase $a^2>b^2>0$ and for the polar phase $a=1, b=0$. The A phase and the polar distorted A phase are chiral and their gap is zero along ${\bf l}={\bf m}\times{\bf n}$, but the polar phase is not chiral and its gap is zero in the plane normal to {\bf m}. All these phases are Equal Spin Pairing (ESP) phases in which the spin susceptibility is the same as in the normal phase and does not depend on temperature.

We identify superfluid phases by measurements of the NMR frequency shift ($\Delta\omega$) from the Larmor value ($\omega_L$). The shift appears due to a dipole interaction of spins of the superfluid condensate and depends on the order parameter, its spatial distribution and $\mu$. In the polar distorted A phase strands of N-aerogel destroy the long-range order in the orbital space: ${\bf m}$ is oriented along $\bf \hat \zeta$, but vectors ${\bf n}$ are uniform only over a scale $\xi_{LIM}\sim1~\mu m$ and at larger distances form a static 2D Larkin-Imry-Ma (LIM) state \cite{we1,we2,Vol} corresponding to a random distribution of ${\bf n}$ in the plane normal to $\bf \hat \zeta$. This state is similar to the LIM state in the A-like phase of $^3$He in silica aerogel \cite{we4, Halp2}.
The vector $\bf d$ is normal to the magnetization $\bf M$ and must be uniform over a dipole length $\xi_d\sim10~\mu m$ determined by the balance between dipole and gradient energies. At larger distances $\bf d$ can be uniform (``spin nematic'', SN, state) or random (``spin glass'', SG, state). The SN state is favorable and usually appears in low excitation NMR experiments. The SG state is metastable and corresponds to a local minimum of the total energy. It can be created in A and polar distorted A phases  by cooling through $T_{ca}$ in presence of high NMR excitation generating a random $\bf d$-distribution \cite{we4}. On further cooling this distribution is ``freezed'' and stabilized by the LIM state. For the 2D LIM state the frequency shift in the SN state is given by the equation \cite{we1,we2}:
\begin{eqnarray}
2\omega_L\Delta\omega=K\Big[\cos\beta-
\frac{\sin^2\mu}{4}\Big(5\cos\beta-1\Big)\Big]\Omega^2_A,
\label{delt}
\end{eqnarray}
where
\begin{equation}
K=\frac{4-6b^2}{3-4a^2b^2},
\label{eqK}
\end{equation}
$\beta$ is the tipping angle of $\bf M$ and $\Omega_A=\Omega_A(T)$ is the Leggett frequency of the A phase (if this phase existed and had the same transition temperature). The Leggett frequency is proportional to the gap and grows from 0 up to $\sim$100~kHz on cooling from the superfluid transition \cite{VW}. For low excitation CW NMR ($\cos\beta\approx$1) and for $\mu=0$ (${\bf H}\parallel \hat\zeta$) we get
\begin{equation}
\label{shift0} 2\omega_L\Delta\omega= K\Omega_A^2,
\end{equation}
while for $\mu=\pi/2$ the shift equals zero. From Eqs.~(\ref{eqK})-(\ref{shift0}) follows that if $\Omega_A$ is known then the value of $\Delta\omega$ for $\mu=0$ allows to determine the value of the polar distortion: in the A phase $K$ should equal 1/2 while for the polar phase $K=4/3$. However, there are two problems. The first problem is that the Leggett frequency was measured only in bulk $^3$He (we denote this value by $\Omega_{A0}$) \cite{la1,la2}. In aerogel $\Omega_A$ is smaller due to suppression of the transition temperature ($\Delta T_{ca}=T_c-T_{ca}$) and the corresponding decrease of the gap. The second problem is that Eq.~(\ref{eqK}) is derived in weak-coupling limit \cite{VW}, which is believed to be a good approximation for bulk $^3$He at low pressures. This equation also does not account for the anisotropy of spin diffusion. Fortunately, in our experiments  $\Delta T_{ca}$ is small (2-10\% of $T_c$ depending on pressure). In this case we can use $\Omega_A$ obtained by rescaling of $\Omega_{A0}$:
\begin{equation}\label{rescale}
 \Omega_A(T/T_{ca})=\frac{T_{ca}}{T_c}\Omega_{A0}(T/T_c).
\end{equation}
Corrections due to the spin diffusion anisotropy are expected to be $\sim$$\Delta T_{ca}/T_c$ \cite{Fom2}, while strong coupling corrections should not exceed $\pm$5\% \cite{Min}. Thus we can conclude that measurements of $\Delta\omega$ for $\mu=0$ allow to estimate the polar distortion with a good accuracy, but in order to distinguish the polar phase from the A phase with a strong polar distortion this is still not enough. However, for this purpose we can compare NMR properties of the SN and SG states.
In the case of low excitation CW NMR the frequency shift in the SG for $\mu=0$ is the same as in the SN state but for $\mu=\pi/2$ the shift in the SG state is negative and equals \cite{we1,we2}:
\begin{equation}
\label{shift90} 2\omega_L\Delta\omega= -CK\Omega_A^2,
\end{equation}
where $C=1/2$ for the isotropic $\bf d$-distribution in the plane normal to $\bf M$. In fact, $C$ may be smaller because for $\mu=\pi/2$ the $\bf d$-distribution becomes anisotropic (vectors $\bf d$ tend to incline towards vectors $\bf n$ which lay in the plane normal to $\hat \zeta$ because it decreases the dipole energy), but the shift remains negative. We note that the order parameter of the polar phase does not contain $\bf n$. It means that the SG state can not be stabilized, i.e. for $\mu=\pi/2$ the shift should always be equal to zero. Thus, the negative shift for $\mu=\pi/2$ indicates that it is the SG state and that the observed phase is not the pure polar phase.
\begin{figure}[t]
\centerline{\includegraphics[width=0.95\columnwidth]{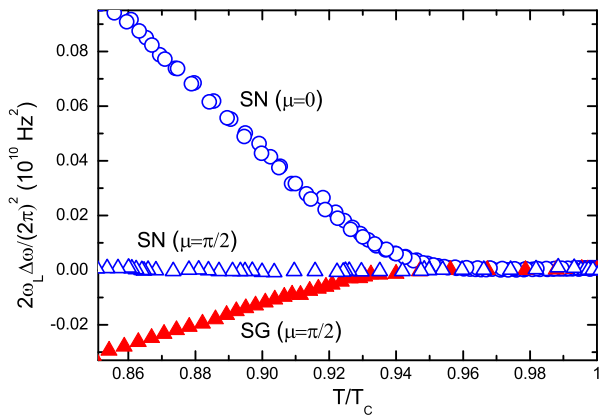}}
\caption{
The frequency shift versus temperature in $^3$He in``Obninsk aerogel'' (Obminsk-50). Open circles -- the SN state for $\mu=0$, open triangles -- the SN state for $\mu=\pi/2$, filled triangles -- the SG state for $\mu=\pi/2$. $T_{ca}\approx 0.94~T_c, P=6.9$~bar. The x-axis represents the temperature normalized to the superfluid transition temperature in bulk $^3$He.}
\label{50mg}
\end{figure}
\begin{figure}[h]
\centerline{\includegraphics[width=0.95\columnwidth]{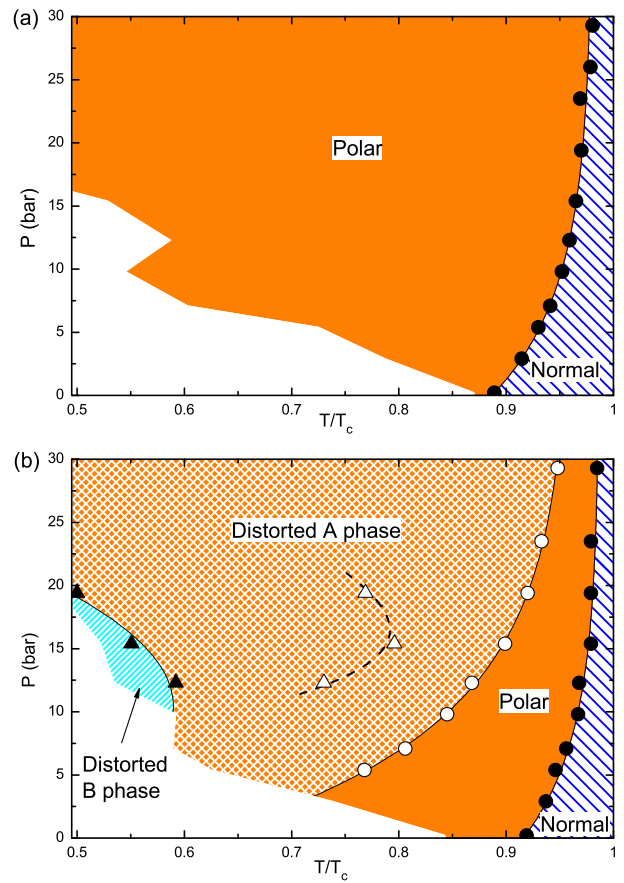}}
\caption{
Phase diagram of $^3$He in nafen-243 (a) and in nafen-90 (b). Filled circles mark the superfluid transition of $^3$He in nafen. Open circles mark the transition between polar and polar distorted A phases. Filled triangles mark the beginning of the transition into the polar distorted B phase on cooling. Open triangles mark the beginning of the transition into the distorted A phase on warming from the distorted B phase. Widths of A-B and B-A transitions are $\sim 0.02\,T_{ca}$. White area shows regions where there are no experimental data.}
\label{pd243}
\end{figure}

{\it Results and Discussion.---}In previous experiments with $^3$He in ``Obninsk aerogel'' with the density of 30~mg/cm$^3$ \cite{we1} it was
found that in the polar distorted A phase the distortion is maximal near $T_{ca}$ where, depending on pressure, values of $K$ were within the range of 0.6$\div$1.07. The maximal value of $K$ was obtained at low pressures and corresponded to $a^2=0.73$ and $b^2=0.27$. This shows that the A phase with a strong polar distortion was definitely obtained, but, taking into account possible corrections, the question about the existence of the pure polar phase remained open. To clarify this, we have done experiments with Obninsk-8 and Obninsk-50 samples. At low pressures near $T_{ca}$ we have obtained that values of $K$ equal $\sim$1.06 and $\sim$1.07 respectively. In contrast to \cite{we1}, we have created the SG state in both samples and found that the shift for $\mu=\pi/2$ is negative and disappears only at $T_{ca}$ (Fig.~2). It points out that the polar phase is still not stabilized in ``Obninsk aerogel''.

The situation is different in $^3$He in nafen, where the polar phase is realized in a wide range of temperatures (see Fig.~3(a,b)). Identification of observed superfluid phases is based on the following:\\
1. The superfluid transition results in the NMR frequency shift for $\mu=0$. The transition occurs into the ESP phase, because the spin susceptibility does not depend on temperature.\\
2. Pulse NMR experiments for different $\mu$ and $\beta$ show that in both samples the spin dynamics in the ESP phase is described by Eq.~(\ref{delt}).\\
3. In nafen-243 $K\approx 4/3$ and practically does not depend on temperature, as it should be in the polar phase \cite{AI} (Fig.~4(a)).\\
4. In nafen-90 $K\approx 4/3$ only in a finite range of temperatures $T_p<T<T_{ca}$ and on further cooling $K$ decreases as expected for the polar distorted A phase (Fig.~4(b)).\\
5. Near $T_{ca}$ values of $K$ are nearly the same in both samples although they have essentially different densities.\\
6. In both samples we were not able to create the SG state by the same procedure which was successful in $^3$He in silica or in ``Obninsk aerogel'': after all of our attempts to create the SG state the
shift for $\mu=\pi/2$ was absent (Fig.~4(a,b)).
\begin{figure}[t]
\centerline{\includegraphics[width=0.95\columnwidth]{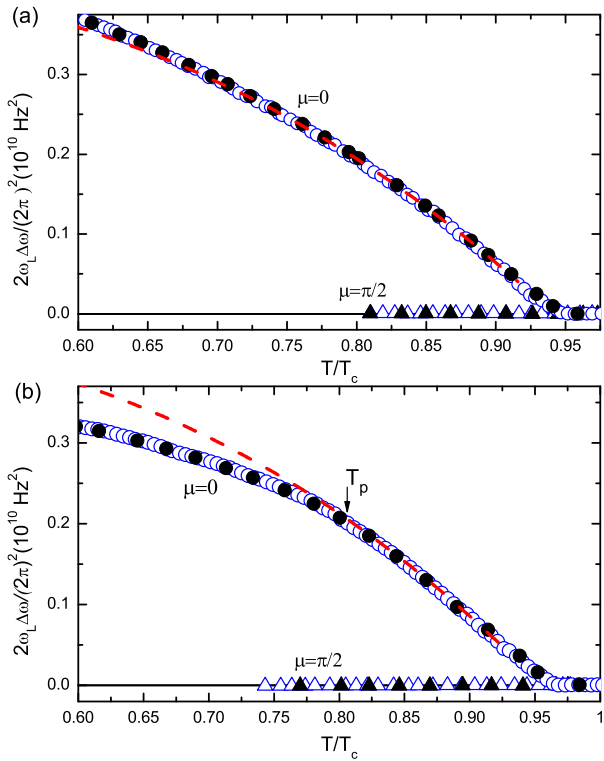}}
\caption{
NMR frequency shifts versus temperature in $^3$He in nafen. Open symbols - the SN state, filled symbols -- data obtained after attempts to create the SG state. $\mu=0$ (circles), $\mu=\pi/2$ (triangles).  $P=7.1$~bar. (a) $^3$He in nafen-243. $T_{ca}\approx 0.94~T_c$. The dashed line corresponds to Eq.~(\ref{shift0}) with $K=1.245$. (b) $^3$He in nafen-90. $T_{ca}\approx 0.955~T_c$. The dashed line corresponds to Eq.~(\ref{shift0}) with $K=1.24$.}
\label{dw243}
\end{figure}
\begin{figure}[t]
\centerline{\includegraphics[width=0.85\columnwidth]{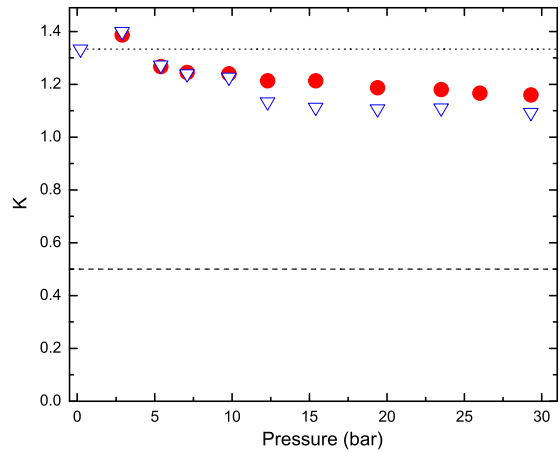}}
\caption{
Values of K measured in the region of existence of the polar phase versus pressure. Open triangles -- $^3$He in nafen-90, filled circles -- $^3$He in nafen-243. Dotted and dashed lines correspond to K expected from Eqs.~(\ref{eqK})-(\ref{shift0}) for pure polar and pure A phases respectively.}
\label{dw90}
\end{figure}

Thus, we assume that in nafen-243 the polar phase exists down to the lowest
attained temperatures, while in nafen-90 it exists only down to $T=T_p$ where the 2nd order
transition into the polar distorted A phase occurs. On further cooling the distortion decreases and experimental values of $\Delta \omega$ deflect from the curve expected for the polar phase.
In Fig.~5 we present the pressure dependence of $K$ determined in the region of existence of the polar phase. At low pressures values of $K$ are close to 4/3, while at high pressures the values are smaller by 10-15\%. This may be due to corrections to Eq.~(\ref{eqK}), which presumably grow with the increase of pressure.

We note that in $^3$He in nafen-90 at low enough temperatures the 1st order transition into the B-like phase occurs which is accompanied by a decrease of the spin susceptibility and by a change of $\Delta\omega$. Properties of this phase were not investigated in detail, but we assume, in analogy with \cite{we3}, that it corresponds to the polar distorted B phase.

In conclusion, we have shown that in $^3$He in nafen the polar phase becomes near $T_{ca}$ more favorable than other possible phases. The polar phase is not stable in bulk $^3$He, but it was obtained in our experiments by using the artificially engineered nanoscale confinement. It is a new topological superfluid and new phenomena can be observed. For example, it is predicted that in this phase half-quantum vortices can be energetically stable \cite{Min}.
Our results also suggest that anisotropic impurities may influence the order parameter in other unconventional superfluids, e.g. in unconventional superconductors or in some of quantum gases.

{\it Acknowledgments.---}We are grateful to I.M.~Grodnensky for providing nafen samples,  R.Sh.~Askhadullin, A.A.~Osipov, P.N.~Martynov for providing samples of ``Obninsk aerogel'', I.A.~Fomin, V.E.~Eltsov and D.E.~Zmeev for useful comments.
This work was supported in part by RFBR (grant 13-02-00674) and by Basic Research Program of the Presidium of Russian Academy of Sciences.


\begin{thebibliography}{99}
\bibitem{VW} D.~Vollhradt and P.~W$\ddot{o}$lfle,  {\it The Superfluid Phases of Helium~3} (Tailor \& Francis, 1990).
\bibitem{ORL} D.D.~Osheroff, W.J.~Gully, R.C.~Richardson, and D.M.~Lee, Phys. Rev. Lett. {\bf 29,} 920 (1972).
\bibitem{Gr73} T.J.~Greytak, R.T.~Johnson, D.N.~Paulson, and J.C.~Wheatley, Phys. Rev. Lett. {\bf 31,} 452 (1973).
\bibitem{a1} W.J.~Gully, D.D.~Osheroff, D.T.~Lawson, R.C.~Richardson, and D.M.~Lee, Phys. Rev. A {\bf 8,} 1633 (1973).
\bibitem{AI} K.~Aoyama and R.~Ikeda, Phys. Rev. B {\bf 73,} 060504 (2006).
\bibitem{Sauls} J.A.~Sauls, Phys. Rev. B  {\bf 88,} 214503 (2013).
\bibitem{Fom} I.A.~Fomin, JETP {\bf 118,} 765 (2014).
\bibitem{Ik} R.~Ikeda, Phys. Rev. B {\bf 91,} 174515 (2015).
\bibitem{Parp} J.V.~Porto and J.M.~Parpia, Phys. Rev. Lett. {\bf 74,} 4667 (1995).
\bibitem{Halp0} D.T.~Sprague, T.M.~Haard, J.B.~Kycia, M.R.~Rand, Y.~Lee, P.J.~Hamot, and W.P.~Halperin, Phys.~Rev.~Lett. {\bf 75,} 661 (1995).
\bibitem{Osh} B.I.~Barker, Y.~Lee, L.~Polukhina, D.D.~Osheroff, L.W.~Hrubesh, and J.F.~Poco, Phys. Rev. Lett. {\bf 85,} 2148 (2000).
\bibitem{weB} V.V.~Dmitriev, V.V.~Zavjalov, D.E.~Zmeev, I.V.~Kosarev, and N.~Mulders, JETP Lett. {\bf 76,} 312 (2002).
\bibitem{we4} V.V.~Dmitriev, D.A.~Krasnikhin, N.~Mulders, A.A.~Senin, G.E.~Volovik, and A.N.~Yudin. JETP Lett. {\bf 91,} 599 (2010).
\bibitem{HalpA} J.~Pollanen, J.I.A.~Li, C.A.~Collett, W.J.~Gannon, and W.P.~Halperin, Phys. Rev. Lett. {\bf 107,} 195301 (2011).
\bibitem{halp15} J.I.A.~Li, A.M.~Zimmerman, J.~Pollanen, C.A.~Collett, and W.P.~Halperin, Phys. Rev. Lett. {\bf 114,} 105302 (2015).
\bibitem{Ask2}  V.E.~Asadchikov, R.Sh.~Askhadullin, V.V.~Volkov, V.V.~Dmitriev, N.K.~Kitaeva, P.N.~Martynov, A.A.~Osipov, A.A.~Senin, A.A.~Soldatov, D.I.~Chekrygina, and A.N.~Yudin, JETP Lett. {\bf 101,} 556 (2015).
\bibitem{we1} R.Sh.~Askhadullin, V.V.~Dmitriev, D.A.~Krasnikhin, P.N.~Martinov, A.A.~Osipov, A.A.~Senin, and A.N.~Yudin, JETP Lett. {\bf 95,} 326 (2012).
\bibitem{we2} R.Sh.~Askhadullin, V.V.~Dmitriev, P.N.~Martynov, A.A.~Osipov, A.A.~Senin, and A.N.~Yudin, JETP Lett. {\bf 100,} 662 (2014).
\bibitem{we3} V.V.~Dmitriev, A.A.~Senin, A.A.~Soldatov, E.V.~Surovtsev, and A.N.~Yudin, JETP {\bf 119,} 1088 (2014).
\bibitem{diff0} R.Sh.~Askhadullin, V.V.~Dmitriev, D.A.~Krasnikhin, P.N.~Martinov, L.A.~Melnikovsky, A.A.~Osipov, A.A.~Senin, and A.N.~Yudin, J. Phys.: Conf. Ser. {\bf 400}, 012002 (2012).
\bibitem{diff} V.V.~Dmitriev, L.A.~Melnikovsky, A.A.~Senin, A.A.~Soldatov, and A.N.~Yudin, JETP Lett. {\bf 101,}(in print)(2015).
\bibitem{Vol} G.E.~Volovik, J. Low Temp. Phys. {\bf 150,} 453 (2008).
\bibitem{Halp2} J.I.A.~Li, J.~Pollanen, A.M.~Zimmerman, C.A.~Collett, W.J.~Gannon, and W.P.~Halperin, Nature Physics {\bf 9,} 775 (2013).
\bibitem{la1} A.I.~Ahonen, M.~Krusius, and M.A.~Paalanen,  J.~Low Temp. Phys. {\bf 25,} 421 (1976).
\bibitem{la2} M.R.~Rand, H.H.~Hensley, J.B.~Kycia, T.M.~Haard, Y.~Lee, P.J.~Hamot, and W.P.~Halperin, Physica B {\bf 194-196,} 805 (1994).
\bibitem{Fom2} I.A.~Fomin, private communication.
\bibitem{Min} V.P.~Mineev, J. Low Temp. Phys. {\bf 177,} 48 (2014).

\end{thebibliography}
\end{document}